\documentclass[12pt]{article}

\usepackage{graphicx}
\usepackage{epstopdf}
\usepackage{amsmath}
\usepackage{amssymb}

\begin{document}

\begin{flushleft}
{\Large
\textbf{A mathematical approach to territorial pattern formation}
}
\\
{\footnotesize
Jonathan R. Potts$^{1,a}$, 
Mark A. Lewis$^{1,2,b}$
\\
\bf{1} Centre for Mathematical Biology, Department of Mathematical and Statistical Sciences, University of Alberta, Canada
\\
\bf{2} Department of Biological Sciences, University of Alberta, Edmonton, Canada
\\
\bf{a} E-mail: jrpotts@ualberta.ca.  Centre for Mathematical Biology, Department of Mathematical and Statistical Sciences, 632 CAB, University of Alberta, Canada, T6G 2G1.  Tel: +1-780-492-1636.

\bf{b} E-mail: mark.lewis@ualberta.ca.  
}
\end{flushleft}

\section*{Abstract}
Territorial behaviour is widespread in the animal kingdom, with creatures seeking to gain parts of space for their exclusive use.  It arises through a complicated interplay of many different behavioural features.  Extracting and quantifying the processes that give rise to territorial patterns requires both mathematical models of movement and interaction mechanisms, together with statistical techniques for rigorously extracting parameters from data.  Here, we give a brisk, pedagogical overview of the techniques so far developed to tackle the problem of territory formation.  We give some examples of what has already been achieved using these techniques, together with pointers as to where we believe the future lies in this area of study.  This progress is a single example of a major aim for 21st century science: to construct quantitatively predictive theory for ecological systems.

\section*{Introduction.}

The natural world is full of complex systems, where constituent parts interact to cause patterns that can be very rich in diversity and unexpected in form.  These range from the detailed termite hills structures that emerge from the collective actions of individually very simple animals, to oscillatory and chaotic patterns in predator-prey systems; from spatially heterogeneous territorial segregations to climactic effects of changing ecosystems \cite{pascual1993, bonabeauetal1998, moorcroftetal2006, breedetal2013}.  To understand how one, more `macroscopic', level of description emerges from finer-grained, `microscopic' processes presents a formidable challenge that can only be tackled with sophisticated mathematical and computational tools, many of which may need to be created for each new problem.

Statistical physics has seen remarkable success at describing emergent phenomena from underlying mechanisms.  Properties that were originally only phenomenologically understood, such as the relationships between heat, pressure and volume of a gas, have been accurately and analytically derived from the movements of tiny, jiggling particles \cite{perrot1998}.  During the 20th century, these ideas have been extended into many areas of physics, including optics, fluid dynamics and soft matter studies, to name but a few \cite{sturge2003}.  However, in the field of physics, the constituent agents are relatively simple particles of inorganic matter.

By contrast, the organisms constituting an ecosystem are living plants and animals, with evolutionarily driven goals and complex behavioral traits.  Despite this complexity, the last few decades have seen many scientists and mathematicians embarking on a journey to develop an analogue of statistical mechanics for ecosystems \cite{levin2012}.  This has been spurred on by an increasing awareness of the need to develop a predictive ecology, so that we can accurately foretell the effects of anthropogenic changes on ecosystems \cite{evansetal2013}.

Here, we review a small part of that journey, the quest to understand how animal populations self-organize into territorial structures from the movements and interactions of individual animals.  Along the way, we will describe a number of mathematical and statistical techniques that have been used to help tackle this problem.  We follow the philosophy that biological questions should drive the decisions to use one particular mathematical technique over another, rather than starting with a field of mathematics and trying to see what it can add to biological understanding.  Consequently, our review is broad rather than deep, intended to entice the reader into reading more about the techniques covered than explain them exhaustively.  We hope that this article will help readers quickly understand the problem of territory formation and introduce them to tools that are helpful in solving problems regarding animal movement, interactions and space use.

\section*{Ecological issues.}  

As well as theoretical curiosity, there are various ecological considerations that make it important to build a mathematical theory of territory formation.  In conservation biology, understanding territory size is vital for designing nature reserves to fit a given population \cite{corsietal1999}.  In epidemiology, if a disease is spreading through a population of territorial animals then it is crucial to understand how the movements and contact rates of the animals relate to the size and shape of the territories \cite{PHG3}.  Territorial interactions can also cause spatial structures to arise that can affect predator-prey dynamics \cite{lewismurray1993}.

Traditionally, territorial structures have been understood by statistical analysis of positional data.  This can often be as simple as drawing the minimum convex polygon around a set of points \cite{harris1990} or assuming that the animal's territory is roughly given by the mean of narrow Gaussian distributions around each observed location \cite{worton1989}.  Recently, more involved techniques that take into account the animal's probable movement between successive locations have been proposed as a more accurate way of determining spatial patterns \cite{benhamou2011}.  It makes sense, therefore, to build on these ideas by also including the interactions between animals into our understanding of territorial formation.  The techniques described in this review explain the mathematical details behind this theory.  Information about the practical lessons provided by such theory is given in our more biologically-oriented companion review \cite{PLPRSB}. 

\section*{Building a model from the ground up: the random walk approach.}

Though animal movement decisions are complex and multi-faceted, we believe the best approach to model building is to start with as simple a model as possible, then build up the complexity one facet at a time, rigorously testing whether each additional term significantly improves how the model fits the available data.  This requires a certain amount of imagination, as a very simple model is unlikely to model well a real animal in an actual environment.  So we start by asking ourselves: what would an animal do if it were placed alone in a barren, featureless landscape?

We might imagine that it moves for a certain time in one direction, to explore.  Then turns at random and moves in another direction for a short while, and so on.  These so-called {\it random walk} patterns have been successfully used as the basis for animal movement models for some time \cite{turchin1998, okubolevin}.  So we start with this idea as the basis for our animal movement models.

\begin{figure}[ht!]
\includegraphics[width=120mm]{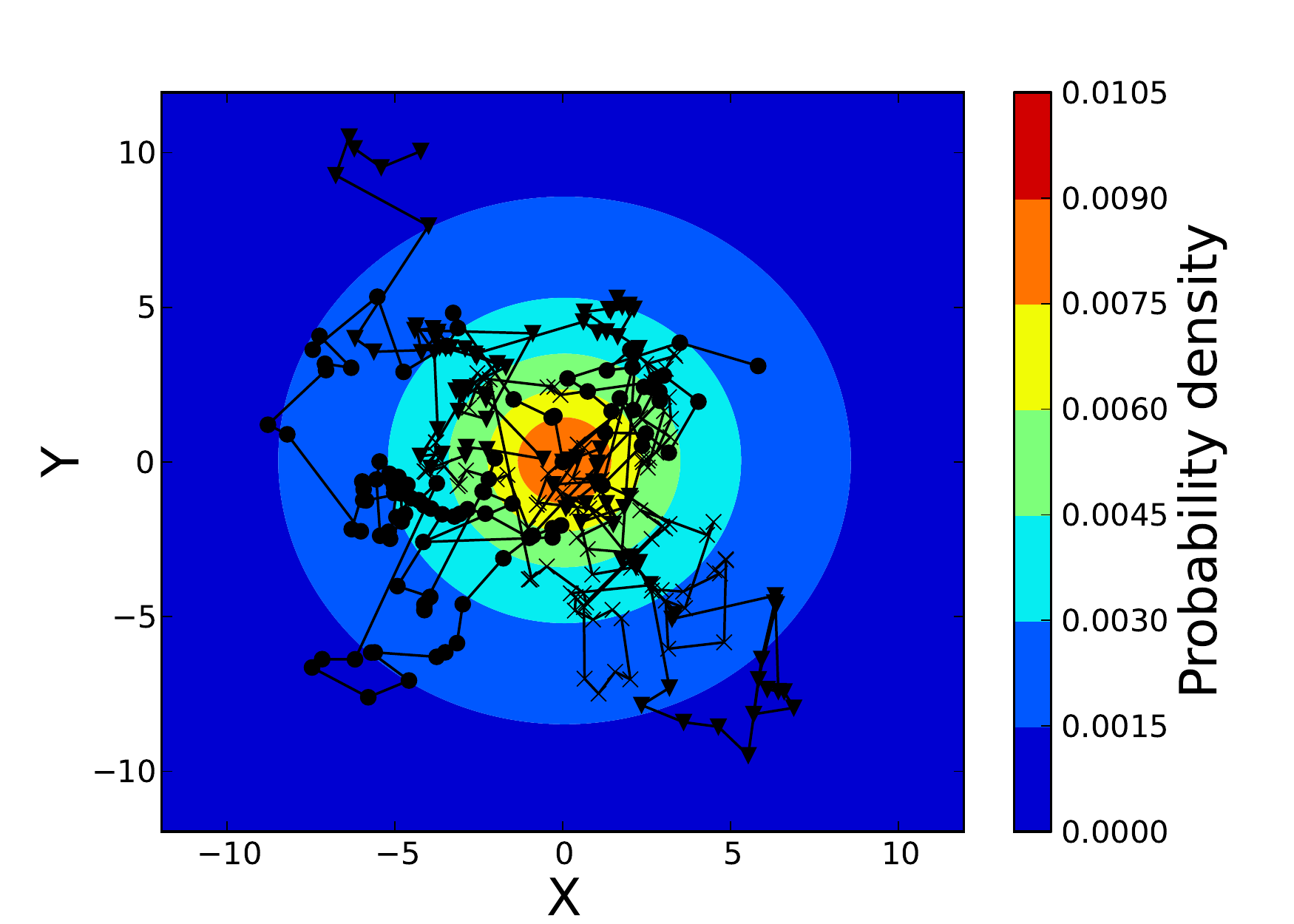}
\caption{\small {\bf Expected spatial patterns from a random walk. }Three paths of 100-step random walks with exponential step-length distribution are shown, as described by equation (\ref{step_length_dist}) with $\delta=1$.  Each path has a different shape, cross, dot or triangle, at the start/end of its steps.  The contours denote the utilisation distribution described by mathematical analysis of the random walk, from equation (\ref{gaussian_ud}).  This displays a very simple example of the predicted spatial patterns that can arise by mathematical analysis of an animal movement model.}
\label{rw_paths}
\end{figure}

Mathematically, a random walk can be described by the probability density $p_\tau({\bf x}|{\bf y})$ of moving to position ${\bf x}$ at time $\tau$ in the future, given that the animal is currently at position ${\bf y}$, which can be in one-, two- or three-dimensional space.  Animals may move a variety of different distances in this time period, but are more likely to move a short distance, and extremely unlikely to move a large distance.  Therefore a suitable distribution of the so-called {\it step lengths}, i.e. lengths of straight-line movement between random turns, might be the exponential distribution:
\begin{align}
p_\tau({\bf x}|{\bf y}) \propto \exp(-\delta|{\bf x}-{\bf y}|),
\label{step_length_dist}
\end{align}
where $\delta$ is a free parameter and the constant of proportionality is calculated by ensuring that the integral of $p_\tau({\bf x}|{\bf y})$ with respect to ${\bf x}$ across the domain of study is 1.  This is often called the {\it step length distribution} of an animal.  Placing such steps together in succession gives a hypothesised possible movement path, as in figure \ref{rw_paths}.

Our ultimate goal is to move from this `microscopic' description of animal movement, to a description of the expected space use of the animal, in a non-speculative, mathematical fashion.  While this is difficult for complex movement models, the random walk equation (\ref{step_length_dist}) is simple enough that we can derive an analytic formula for the space use patterns.  We show this here in the one-dimensional (1D) case.

Let $u(x,t)$ be the probability that the animal is in position $x$ at time $t$. (We no longer use bold font since the positions are 1D.)  Then we can write down the so-called {\it master equation} for this system as 
\begin{align}
u(x,t+\tau) = \int_{-\infty}^{\infty}p_\tau(x|y)u(y,t){\rm d}y.
\label{me}
\end{align}
Given some initial distribution $u(x,0)=u_0(x)$, this can be solved numerically over time.  Moreover, by taking the limit as $\tau \rightarrow 0$, it is possible to derive a diffusion equation, which is a type of partial differential equation (PDE).  

To do this, we first define $z=y-x$ to give the following master equation 
\begin{align}
u(x,t+\tau) = \frac{\delta}{2}\int_{-\infty}^{\infty} \exp(-\delta |z|)u(x+z,t){\rm d}z.
\label{me2}
\end{align}
A Taylor expansion of $u(x+z,t)$ gives
\begin{align}
u(x,t+\tau) = \frac{\delta}{2}\int_{-\infty}^{\infty} \exp(-\delta |z|)\biggl[ & u(x,t)+z\frac{{\partial}u}{{\partial}z}(x,t)+ \nonumber \\ &\frac{z^2}{2}\frac{{\partial}^2u}{{\partial}z^2}(x,t)+O(z^3)\biggr]{\rm d}z.
\label{me3}
\end{align}
Rearranging this, and calculating the integrals of $z\exp(-\delta |z|)$ and $z^2\exp(-\delta |z|)$ gives
\begin{align}
u(x,t+\tau) - u(x,t) = \frac{1}{2\delta^2}\frac{{\partial}^2u}{{\partial}x^2}(x,t)+\frac{\delta}{2}\int_{-\infty}^{\infty}\exp(-\delta |z|)O(z^3){\rm d}z.
\label{me4}
\end{align}
Now we need to take the limit as $\tau \rightarrow 0$.  When doing this, we need to notice that $\delta$ will not remain constant, because the animal's step length distribution will narrow as the time intervals become smaller.  

The only way to obtain a sensible limit for equation (\ref{me4}), i.e. where $\partial u/\partial t$ is not zero or infinity, is to insist that $\tau$ goes to $0$ in such a way that $D = 1/2\delta^2\tau$ is kept constant.  Then 
\begin{align}
\frac{u(x,t+\tau) - u(x,t)}{\tau} = D\frac{{\partial}^2u}{{\partial}x^2}(x,t)+O(\sqrt{\tau}),
\label{me5}
\end{align}
where the $O(\sqrt{\tau})$ arises from the fact that the $n$-th order moment of the exponential distribution is propotional to $\delta^{-n}$, which scales as $\tau^{-n/2}$ in the limit.  Taking the limit as $\tau \rightarrow 0$ gives
\begin{align}
\frac{\partial u}{\partial t} = D\frac{{\partial}^2u}{{\partial}x^2}(x,t),
\label{diffneq}
\end{align}
which is the classical diffusion equation.  

Since the animal starts in a given position $x_0$, the initial condition is the Dirac delta function $u_0(x)=\delta(x_0)$.  The exact solution of equation (\ref{diffneq}), with this initial condition, is just a normal distribution with variance that increases linearly in time 
\begin{align}
u(x,t)=\frac{\exp[-(x-x_0)^2/2Dt]}{2\sqrt{\pi Dt}}.
\label{gaussian}
\end{align}
This result, in fact, generalises to higher dimensions.  In the case of animal territoriality, we are interested in the 2D result
\begin{align}
u({\bf x},t)=\frac{\exp[-({\bf x}-{\bf x}_0)^2/2Dt]}{4\pi Dt},
\label{gaussian2d}
\end{align}
where the power of two denotes the scalar product of a vector with itself.  

This gives our first example of the sort of space use patterns that can be shown to arise from descriptions of animal movement.  If the animal moves in a random fashion, its space use distribution at any point in time is described by equation (\ref{gaussian2d}).  Of course, animal movement is typically far from random.  We explain how to add realism in the next section.

When estimating space use patterns in reality, field biologists have to measure positions over a period of time.  As such, they cannot derive the probability distribution at any particular point in time, but instead construct a utilisation distribution, sometimes called a {\it home range} \cite{burt1943}, over a given time window such as a day, month or season.  To compare this simple random walk model with the positional data, it is necessary to make use of the utilisation distribution derived from equation (\ref{gaussian2d}) over the time $T$ during which the data were collected (figure \ref{rw_paths})
\begin{align}
\frac{\int_0^{T}u({\bf x},t){\rm d}t}{\int_\Omega \left[\int_0^{T} u({\bf x},t){\rm d}t\right]{\rm d}{\bf x}}.
\label{gaussian_ud}
\end{align}

\section*{Getting the model right: statistical techniques.}

Real animals, of course, do not exist alone in featureless environments.  The natural drive to survive and reproduce means that interactions with the environment and other animals greatly affect movement decisions.  For example, the needs to gather food, avoid predation, maintain territories, or find mates may all be contributing factors in an animal's movement choices \cite{fortinetal2005, latombeetal2013, vanakaetal2013}.  

The challenge is both to disentangle which of these factors make significant contributions to movement, and to quantify their effects.  By doing this, we can construct realistic models of animal movement, from which accurate predictions of spatial patterns should emerge.  

There are two main approaches to constructing such models in a rigorous, data-driven fashion.  One is the {\it hypothesis testing} approach, whereby parameters are added to the model one at a time.  Each time a parameter is added, we test whether it significantly improves the model.  If so, we keep it; if not, it is discarded.  The second is the {\it model selection} approach, whereby a number of plausible models are constructed and fitted to the data.  We then use the best one to construct our model of space use.

In the context of animal movement, the so-called {\it likelihood ratio test} provides a useful means for hypothesis testing.  Suppose we have a simple model of animal movement, for example the random walk model of equation (\ref{step_length_dist}).  Denote our data on the positions of the animal at times $0,\tau,2\tau,\dots,N\tau$ by ${\bf x}_0,{\bf x}_1,\dots,{\bf x}_N$.  

Since we have spent many weeks observing the animals, or have spent many hours listening carefully to someone who has, we have sufficient intuition to construct hypotheses about the animals' movements.  For example, we might hypothesise that they have a tendency to move towards areas of high resource biomass.  If $b({\bf x})$ is the resource biomass at position ${\bf x}$ then we could construct the following model to take this into account
\begin{align}
p^1_\tau({\bf x}|{\bf y},\alpha) \propto \exp[-\delta |{\bf x}-{\bf y}|+\alpha b({\bf x})],
\label{resource_model}
\end{align}
which is the probability of moving from ${\bf y}$ to ${\bf x}$ in a time-interval $\tau$, analogous to equation (\ref{step_length_dist}).  This model assumes implicitly that the distribution of step lengths, disregarding the effect of resources, is exponential.  It also assumes that the effect of resources on movement is proportional to the exponential of the biomass.  Neither of these are necessarily true for a given data set, so it is important in practice to try several functional forms for equation (\ref{step_length_dist}) and use a model selection method (e.g. the Akaike Information Criterion detailed in the next section) to find the best one.  We are just using the model in equation (\ref{step_length_dist}) for ease of explanation and mathematical exposition.  

The null hypothesis $H_0$ is that $\alpha=0$, hypothesising that the best model is equation (\ref{step_length_dist}), whereas the alternative hypothesis $H_1$ is that $\alpha \neq 0$.  Testing this requires constructing the so-called {\it likelihood function}, which is the probability of the data given the model and parameters. If we assume that the movement steps are independent from one another (which may not always be true, but is a surpisingly fruitful assumption \cite{PHG3, fortinetal2005, latombeetal2013, vanakaetal2013}), the likelihood takes the following form
\begin{align}
L({\bf x}_0,\dots,{\bf x}_N|\alpha)=\prod_{n=0}^{N-1}p^1_\tau({\bf x}_{n+1}|{\bf x}_n,\alpha).
\label{likelihood_fn_1}
\end{align}
The method of maximum likelihood chooses the parameters so as to make the likelihood as large as possible. Let $\alpha_{\rm max}$ be the value of $\alpha$ that maximises expression (\ref{likelihood_fn_1}).  Then the {\it likelihood ratio test} tells us that $2\log[L({\bf x}_0,\dots,{\bf x}_N|\alpha_{\rm max})]-2\log[L({\bf x}_0,\dots,{\bf x}_N|0)]$ is approximately $\chi$-squared distributed with 1 degree of freedom \cite{burnhamandersen}.  

\begin{figure}[ht!]
\centering
\includegraphics[width=120mm]{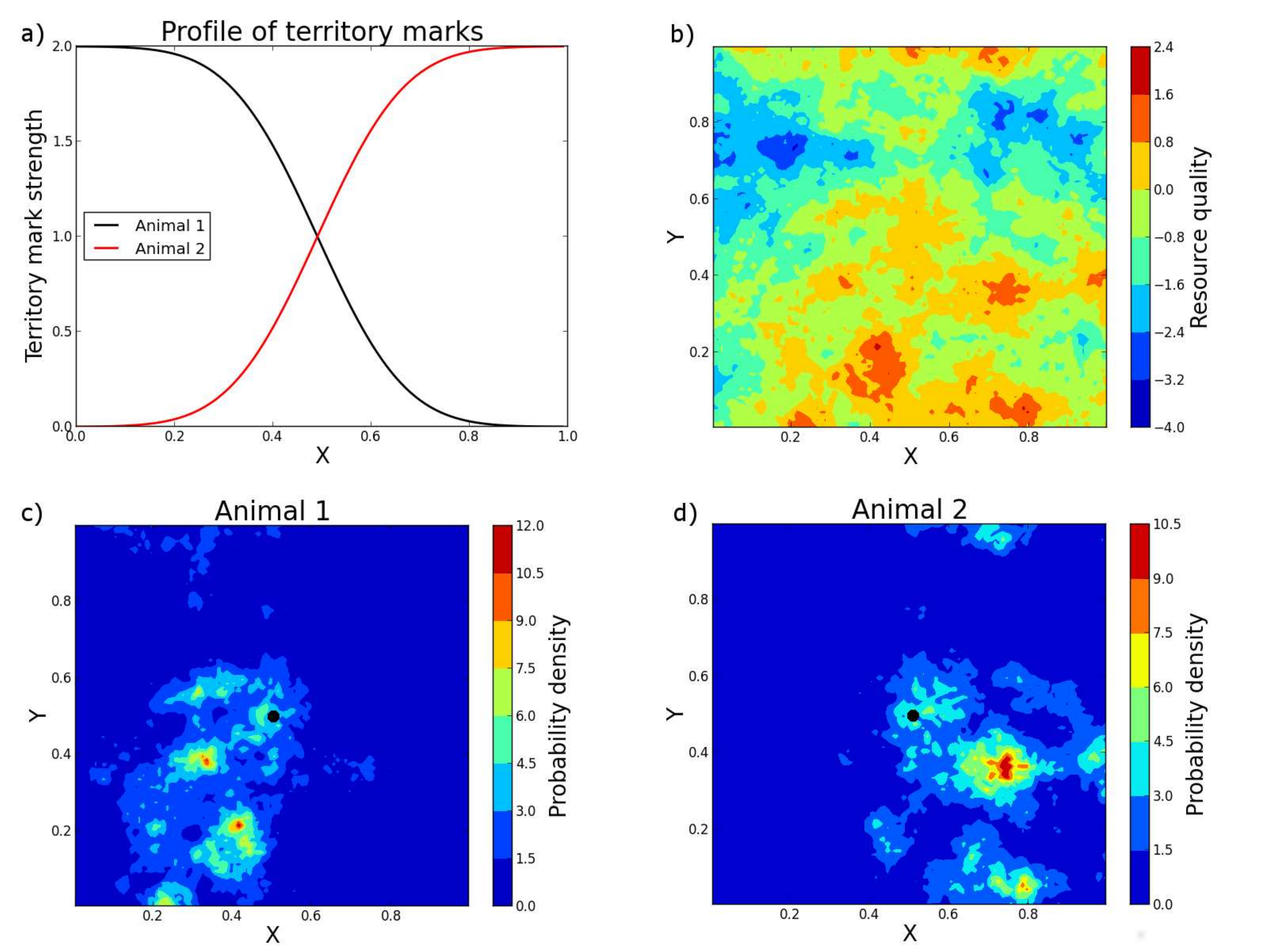}
\caption{{\bf Where next?}  This example demonstrates the probability of an animal's next move, dependent on two factors: (a) collective interactions via the strength of conspecific territorial marks and (b) environmental interactions via resource quality.  Both aspects can be modelled separately or together, then the models can be tested against the data using hypothesis testing or model prediction to find out if either significantly affect movement processes.  The strength of territory marks in this example does not change in the $Y$-direction, so that animal 1 has territory on the left and animal 2 on the right.  The probability of animal 1's (resp. animal 2's) next position after some time interval $\tau$, given that it's current position is in the middle of the landscape (black dot), is shown in panel (c) (resp, panel d).  As each animals moves, it marks the terrain causing the territorial profile to change over time, which in turn influences the other animal's movements.  This feedback mechanism can cause territorial confinement to emerge (reproduced from \cite{PML}).}
\label{cssf}
\end{figure}

We can therefore use the $\chi$-squared test to test whether there is sufficient evidence to reject $H_0$.  If we reject $H_0$ then we can consider $p^1_\tau({\bf x}|{\bf y},\alpha_{\rm max})$ as an improved model of animal movement, compared to the simple random walk model of equation (\ref{step_length_dist}).  We may therefore use it to compute predicted space use patterns, via constructing the master equation as per equation (\ref{me}), and if possible taking a PDE limit (see \cite{moorcroftbarnett2008, barnettmoorcroft2008} for an example).  The predictions can then be compared with those of the diffusion equation (\ref{gaussian2d}) to see if they are more accurate.  Proceeding in this way, we can add parameters one at a time to improve the fit of our model to the data until we are satisfied with the space use predictions (figure \ref{cssf}).

As an alternative to the reductionist approach of hypothesis testing, we may wish to use the approach of multiple working hypotheses.  Here we would formulate an array of plausible models, each defined by a different group of nonzero parameters, and see which of these models is best supported by the data. This can be more computationally intensive, but it takes into account the idea that a blend of different factors may affect movement, and that the effects might only be observed when all the covariates are included at the same time.  

As a simple example, suppose that both resource biomass and the distance $|{\bf x}-{\bf x}_c|$ from a predator's home range centre, ${\bf x}_c$, are hypothesised to influence movement decisions.  Then we can construct four different movement models
\begin{align}
p^0_\tau({\bf x}|{\bf y},\alpha) &\propto \exp[-\delta |{\bf x}-{\bf y}|], \nonumber \\
p^1_\tau({\bf x}|{\bf y},\alpha) &\propto \exp[-\delta |{\bf x}-{\bf y}|+\alpha b({\bf x})], \nonumber \\
p^2_\tau({\bf x}|{\bf y},\alpha) &\propto \exp[-\delta |{\bf x}-{\bf y}|+\beta |{\bf x}-{\bf x}_c|], \nonumber \\
p^3_\tau({\bf x}|{\bf y},\alpha) &\propto \exp[-\delta |{\bf x}-{\bf y}|+\alpha b({\bf x})+\beta |{\bf x}-{\bf x}_c|].
\label{aic_models}
\end{align}
The Akaike information criterion (AIC) gives a technique for choosing between these models.  The AIC of a model is $2k-2\log(L_{\rm max})$ where $L_{\rm max}$ is the maximum of the likelihood function and $k$ the number of model parameters.  Intuitively, it measures the relative closeness of models to the data, with some penalisation for models with a larger number of parameters, although the exact form of the AIC expression can be derived more deeply using information theory \cite{burnhamandersen}.  The various models can be used to build a mechanistic model of animal movement, by constructing the master equation as in equation (\ref{me}).  The model with the lowest AIC is likely to describe the space use patterns most accurately.

Once we have our best model, though, it is important to ask how much better it is than the rest.  In other words, what is the chance that we are wrong and one of the other models is in fact the best?  The theory of AIC gives a nice answer for this.  Let $\mbox{AIC}_{\rm min}$ denote the AIC of the best model and $\mbox{AIC}_{\rm i}$ that of model $i$.  Then each model $i$ is $\exp[(\mbox{AIC}_{\rm min}-\mbox{AIC}_{\rm i})/2]$ times as likely as the model with the minimum AIC to be the `best' model, insofar as it minimises information loss (see \cite{burnhamandersen}).  

We have given something of a whistle-stop tour of model selection and hypothesis testing, emphasising the relations to building animal movement models.  There are a number of textbooks that give detailed descriptions of AIC, likelihood and related topics, e.g. \cite{burnhamandersen}, to which we refer the interested reader for more information.  For an example of model selection and hypothesis testing in the context of movement models, see \cite{PMSL}.  Sometimes authors use AICc, Bayesian Informaion Criteriion (BIC) or other related techniques, which have various pros and cons, which are discussed in detail elsewhere (e.g. \cite{burnhamandersen}).

\section*{Adding territorial interactions.}

Now we have the tools to build up a model of individual animal movement, we come to the main aspect of this paper: accounting for territorial interactions.  There are two approaches to this in the modelling literature.  The first, typified in the book by \cite{moorcroftlewis2006}, is to derive space use patterns from a plausible model of interaction mechanisms, then fit these patterns to location data.  One can then test various candidate models against the data, for example using the AIC techniques described in the previous section, to infer information about the drivers of territorial structures.  

The second is to continue along a similar path as in the previous section, fitting the individual movement and interaction model to data, then see whether the spatial patterns that emerge are similar to the territories described by animal locations.  This more conservative approach is newer and less well-developed, but as such raises a number of interesting challenges for future developers of territorial models.

Perhaps the first use of partial differential equations to capture territories emerging from animal movements and interactions was that of \cite{lewismurray1993}, in the context of modelling wolf pack territoriality.  The simplest model in that paper was of two packs with densities $U(x,t)$ and $V(x,t)$ and can be derived on a 1D lattice with zero-flux boundary conditions \cite[section 3.2]{moorcroftthesis}.  It posited that as wolves move, they deposit scent which decays at a rate $\mu$.  The density of scent for packs $U$ and $V$ at position $x$ and time $t$ are denoted $P(x,t)$ and $Q(x,t)$ respectively.  The rate of scent deposition of pack $U$ is $l+\nu Q(x,t)$, and is $l+\nu P(x,t)$ for pack $V$, modelling the fact that there is a baseline rate of scent deposition, but that it also increases in the presence of the other pack's scent.  

The wolves have a constant speed but switch direction at a rate dependent on the presence of conspecific scent.  We assume that the den site of pack $U$ is at the left-hand boundary of the lattice, while pack $V$ has den site to the right.  Therefore the probability of a member of pack $U$ switching from right to left (resp. left to right) is $\lambda+\sigma Q(x,t)$ (resp. $\lambda-\sigma Q(x,t)$), whereas the probability of a member of pack $V$ switching from left to right (resp. right to left) is $\lambda+\sigma P(x,t)$ (resp. $\lambda-\sigma P(x,t)$).  Letting the positions of left- and right-moving members of pack $U$ be denoted by $U^-(x,t)$ and $U^+(x,t)$ respectively, and similarly for $V$, we arrive at the following master equation in discrete space and time \cite[section 3.2]{moorcroftthesis}
\begin{align}
U^-(x,t+\tau)=&[1-\tau\{\lambda-\sigma Q(x,t)\}]U^-(x+a,t)+ \nonumber \\ 
&\tau[\lambda+\sigma Q(x,t)]U^+(x-a,t), \nonumber \\
U^+(x,t+\tau)=&[1-\tau\{\lambda+\sigma Q(x,t)\}]U^+(x-a,t)+ \nonumber \\
&\tau[\lambda-\sigma Q(x,t)]U^-(x+a,t), \nonumber \\
V^-(x,t+\tau)=&[1-\tau\{\lambda-\sigma P(x,t)\}]V^-(x+a,t)+ \nonumber \\ 
&\tau[\lambda+\sigma P(x,t)]V^+(x-a,t), \nonumber \\
V^+(x,t+\tau)=&[1-\tau\{\lambda+\sigma P(x,t)\}]V^+(x-a,t)+ \nonumber \\
&\tau[\lambda-\sigma P(x,t)]V^-(x+a,t), \nonumber \\
P(x,t+\tau)=&(1-\mu\tau)P(x,t)+U(x,t)[l+\nu Q(x,t)]\tau, \nonumber \\
Q(x,t+\tau)=&(1-\mu\tau)Q(x,t)+V(x,t)[l+\nu P(x,t)]\tau,
\label{lewis_terr_me}
\end{align}
where $a$ is the lattice spacing and $\tau$ the time it takes to move distance $a$.  

To derive partial differential equations from this stochastic model, a {\em mean field} approximation is needed that assumes the distributions of scent marks and individuals are uncorrelated. This could be unreasonable on short time scales, but is reasonable on the longer times scales that are relevant to the formation of territorial patterns since the movement of the individuals is rapid compared to the change in scent mark density. Again, we need to take a delicate limiting process.  Based on our experience with the earlier diffusion limit taken on equation (\ref{me4}), and observing that the space step $a$ takes the place of the mean space step $\delta^{-1}$ in equation (\ref{me2}), we would expect that the space and time steps, $a$ and $\tau$ would approach zero with $a$ scaling so that $a^2/\tau$ is constant.  However, we now have additional parameters, $\lambda$ and $\sigma$. How should they scale?  
A natural choice is to have $\lambda $ increase so that $\lambda \tau$ approaches a constant as $\tau$ approaches zero, and to have $\sigma$ increase so that $\sigma a$
approaches a constant as $a$ approaches zero. That way, the switching and scent mark bias terms are incorporated into model as significant factors during the limiting process. Denoting by lower case letters the densities that correspond to upper case letters for probabilities in equation (\ref{lewis_terr_me}), we arrive at the following PDEs describing the emergent space use patterns (see \cite[section 3.2]{moorcroftthesis} for a derivation)
\begin{align}
\frac{\partial u(x,t)}{\partial t} -&c\frac{\partial}{\partial x}[q(x,t)u(x,t)]= d\frac{\partial^2 u(x,t)}{\partial x^2},\nonumber \\
\frac{\partial v(x,t)}{\partial t} +&c\frac{\partial}{\partial x}[p(x,t)v(x,t)]= d\frac{\partial^2 v(x,t)}{\partial x^2},\nonumber \\
\frac{\partial p(x,t)}{\partial t} =& [l+\nu q(x,t)]u(x,t) - \mu p(x,t), \nonumber \\
\frac{\partial q(x,t)}{\partial t} =& [l+\nu p(x,t)]v(x,t) - \mu q(x,t),
\label{lewis_terr_pde}
\end{align}
where $(\sigma a)/(\tau \lambda)\rightarrow c$ and $(a^2)/(2\tau^2 \lambda)\rightarrow d$ as $a$ and $\tau$ tend to zero.  

It is interesting to consider other possible limiting processes.  For example, we could argue, quite reasonably, that the speed of movement $a/\tau$ should remain constant during the limiting process rather than becoming arbitrarily large.  This limit is mathematically possible, and leads to related hyperbolic models for animal movement (see \cite{hillenpainter2013} for the general theory).  However, as argued in \cite{moorcroftthesis}, the hyperbolic and parabolic limiting equations have similar behaviour when evaluating territorial pattern formation over long time scales.

\begin{figure}[ht!]
\centering
\includegraphics[width=120mm]{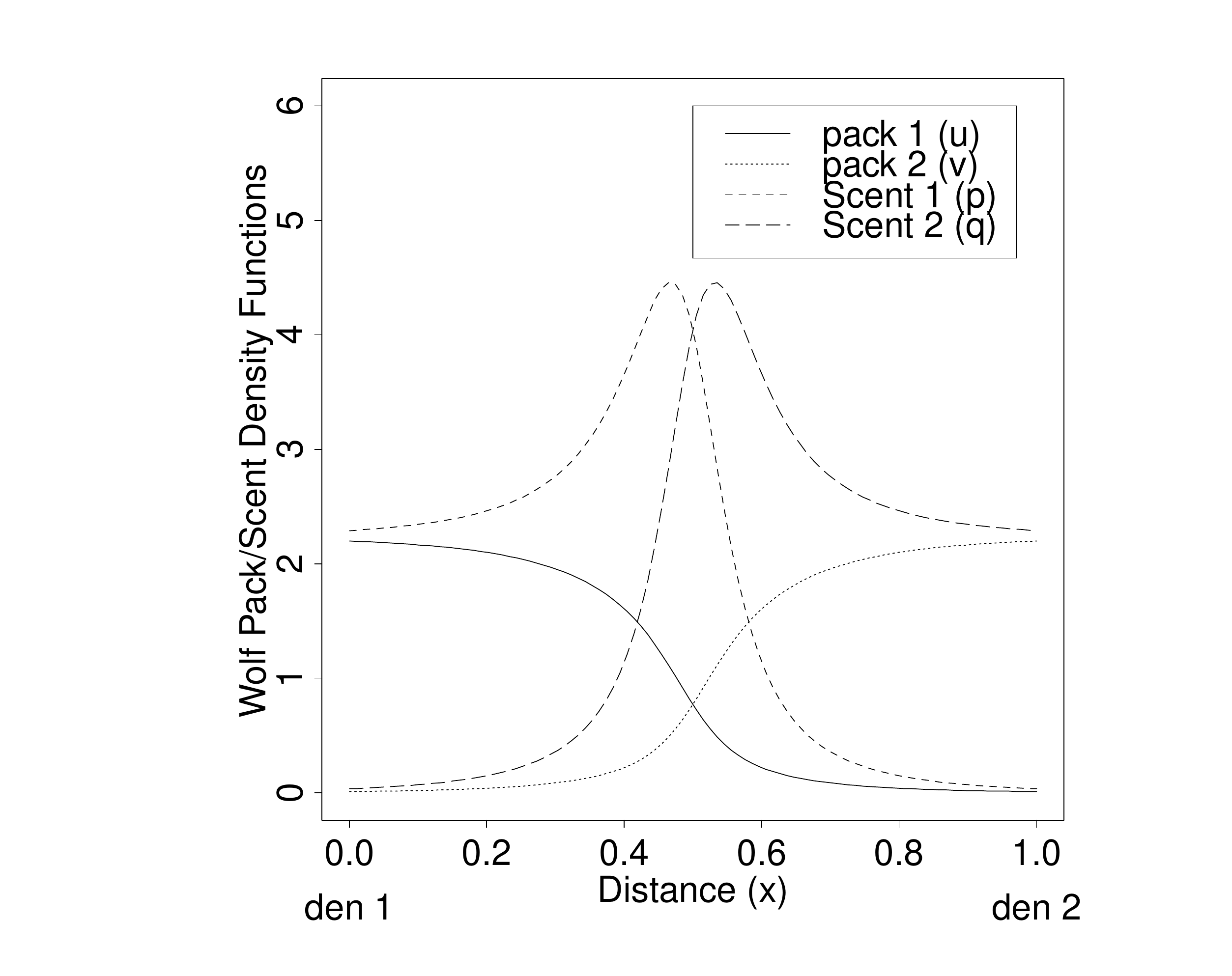}
\caption{{\bf One-dimensional model results}  
Sample solutions for the one-dimensional partial differential equation model {\protect equation (\ref{lewis_terr_pde})}. Note the segregation of $u$ and $v$ and the bowl-shaped scent densities for $p$ and $q$ (based on {\protect \cite{lewis1997}}).}
\label{1dmodel}
\end{figure}

Thus we have a system of PDEs describing territorial patterns that is rigorously derived from the underlying movement and interaction processes.  These equations generate steady-state solutions that exhibit spatial patterns that correspond qualitatively to found in territories \cite{lewis1997}  (figure \ref{1dmodel}). Here there is spatial segregation between the  the two packs, $u$ and $v$, and the scent marks, $p$ and $q$ are highest along the boundary between the two packs. To understand this qualitatively we see that segregation arises from advection terms in equation (\ref{lewis_terr_pde}) that drive individuals back towards their den site when the encounter foreign scent marks, and heightened scent mark densities at boundaries arise from a positive feedback loop where scent marking from one pack gives rise to heightened scent marking rates from the other pack. Indeed, it actually possible to choose a feedback loop that is so strong that scent marks exhibit mathematical ``blow up'' at territorial boundaries \cite{lewis1997}. This is  intriguing, although biologically nonsensical, if only due to the finite bladder capacity of animals involved.

Although interesting to analyse, these equations are only described in 1D and contain no behavioural features other that scent marking and conspecific avoidance.  To add further realism, it is necessary first to extend the results into 2D, then add further plausible drivers of spatial patterns to the PDE, to create a suite of possible models that describe predicted territorial distributions.  These distributions can be fitted to data on animal locations to deduce which model is the best at explaining the complex patterns observed in nature.

We demonstrate this with an example from \cite{moorcroftetal2006}, where the authors use a mechanistic model of territory formation to determine the movement tendencies that underlie coyote territories.  They start with a 2D version of the model in equations (\ref{lewis_terr_pde}) that applies to $n$ packs with position densities $u_1({\bf x},t),\dots,u_n({\bf x},t)$ and scent densities $p_1({\bf x},t),\dots,p_n({\bf x},t)$ 
\begin{align}
\frac{\partial u_i({\bf x},t)}{\partial t} =& d\nabla^2 u_i({\bf x},t)-c\nabla \cdot \left[{\bf x}_iu_i({\bf x},t)\sum_{j\neq i}q_j({\bf x},t)\right],\nonumber \\
\frac{\partial p_i({\bf x},t)}{\partial t} =& \left[l+\sum_{j\neq i}\nu p_j({\bf x},t)\right]u_j({\bf x},t) - \mu p_i({\bf x},t),
\label{lewis_terr_2Dpde}
\end{align}
where ${\bf x}_i$ is the unit vector in the direction from ${\bf x}$ to the den site for pack $i$.  This equation can be derived from a biased random walk process with scent-marking in two spatial dimensions (see \cite{moorcroftlewis2006} for details). By non-dimensionalising appropriately \cite{moorcroftlewis2006} and assuming that the system is contained within a finite domain with zero flux boundary conditions, the system steady state is given by
\begin{align}
0 =& \nabla^2 u_i({\bf x})-\beta\nabla \cdot \left[{\bf x}_iu_i({\bf x})\sum_{j\neq i}q_j({\bf x})\right],\nonumber \\
0 =& \left[1+\sum_{j\neq i}m p_j({\bf x})\right]u_j({\bf x},t) - p_i({\bf x}). 
\label{lewis_terr_2Dpde_ss}
\end{align}
To this system of ordinary differential equations (ODEs), the authors add two different terms.  The first corresponds to a tendency for coyotes to move away from steep terrain, where it's difficult for them to roam, giving the following model
\begin{align}
0 =& \nabla^2 u_i({\bf x})-\beta\nabla \cdot \left[{\bf x}_iu_i({\bf x})\sum_{j\neq i}q_j({\bf x})\right] - \nabla [\alpha_z u_i({\bf x})\nabla z({\bf x})],
\label{lewis_terr_2Dpde_sta}
\end{align}
where $z({\bf x})$ is the elevation of the landscape and $\alpha_z$ the strength of the tendency to move away from high ground.  The second models the tendency to move towards areas with higher prey availability, giving the following model
\begin{align}
0 =& \nabla \cdot [{\rm e}^{-\alpha_rh({\bf x})}\nabla u_i({\bf x})]-\beta\nabla \cdot \left[{\rm e}^{-\alpha_rh({\bf x})}{\bf x}_iu_i({\bf x})\sum_{j\neq i}q_j({\bf x})\right] - \nonumber \\
& \nabla \cdot [{\rm e}^{-\alpha_rh({\bf x})}u_i({\bf x})\nabla h({\bf x})],
\label{lewis_terr_2Dpde_pa}
\end{align}
where $h({\bf x})$ is the amount of prey available and $\alpha_r$ a measure of the strength of the tendency to move towards areas of higher prey availability.  The three models in equations (\ref{lewis_terr_2Dpde_ss}), (\ref{lewis_terr_2Dpde_sta}) and (\ref{lewis_terr_2Dpde_pa}) are then fitted to data on coyote locations to determine which gave the best fit and therefore which gives the most likely explanation as to the causes of territorial patterns.

\begin{figure}[ht!]
\includegraphics[width=\columnwidth]{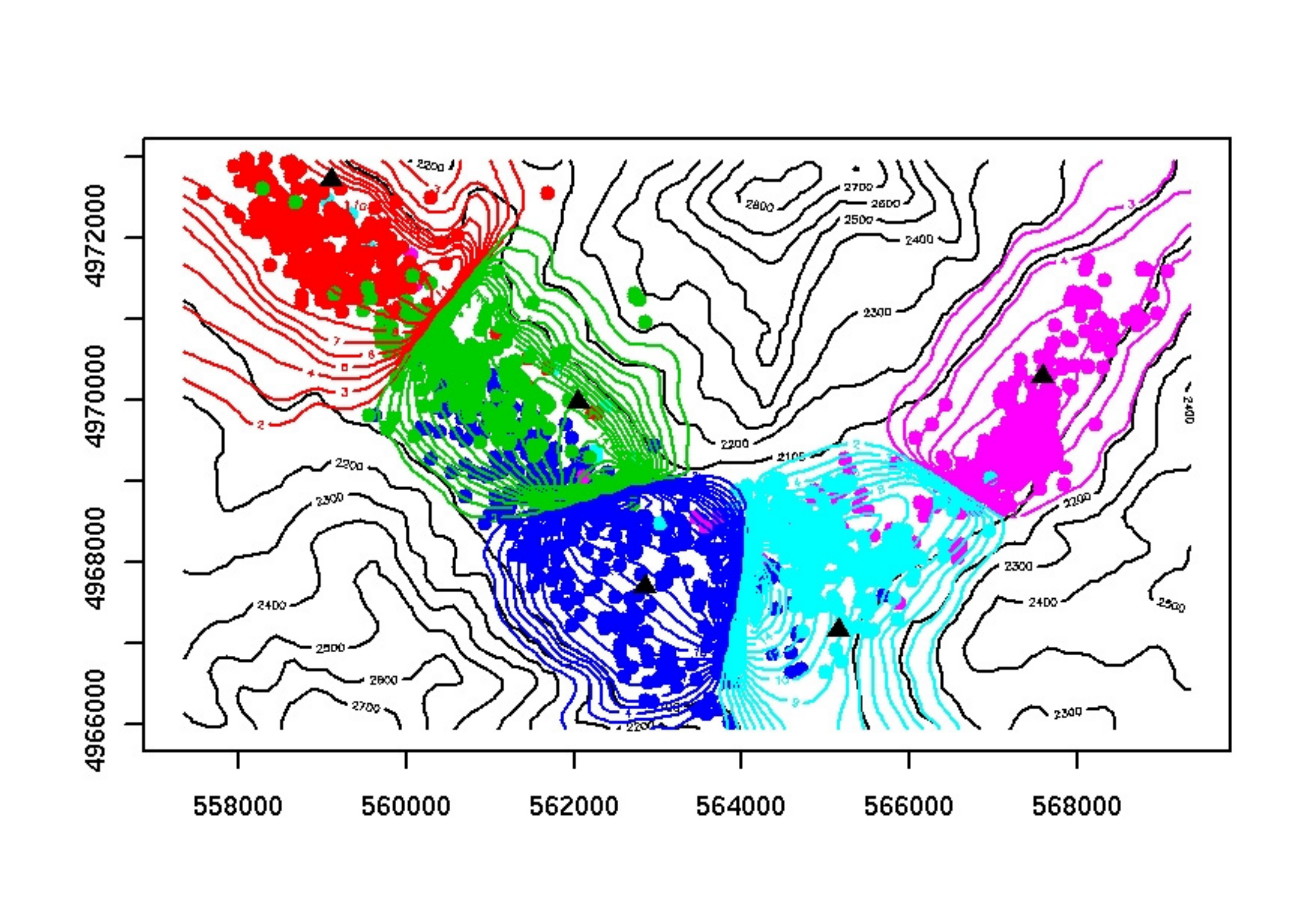}
\includegraphics[width=\columnwidth]{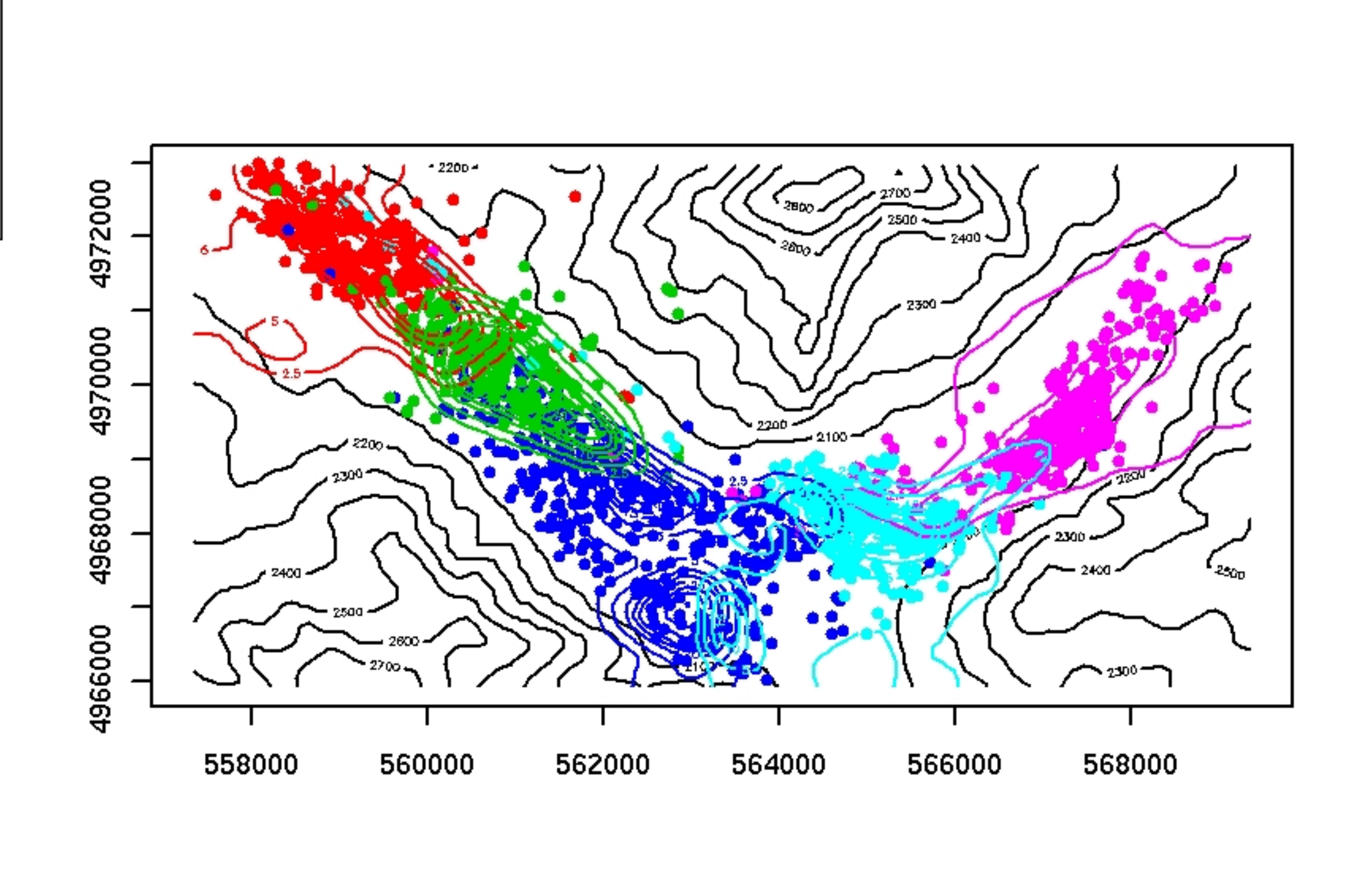}
\caption{\small {\bf Mechanistic models capturing coyote territorial patterns.}  The top panel shows the best fit of equation (\ref{lewis_terr_2Dpde_sta}) to data on coyotes in Lamar Valley, Yellowstone National Park and the bottom panel shows the same for equation (\ref{lewis_terr_2Dpde_pa}).  Contour lines show the space use distributions of the best-fit model, whereas dots give relocation fixes for coyotes.  Different colours represent positions of different packs. The coordinates are measure in UTM. Reproduced with permission from {\protect \cite{moorcroftetal2006}}.}
\label{MLMplot}
\end{figure}

The procedure for fitting to data uses an AIC test, by taking locations sufficiently far apart (in both space and time) so that they can be considered independent random variables drawn from the steady state distribution.  For each pack $i$, let ${\bf x}_0,\dots,{\bf x}_{N_i}$ be the positions of these independent relocations.  Then the likelihood of the data given the model is
\begin{align}
L=\prod_{i=1}^n\prod_{{k_i}=0}^{N_i}u_i({\bf x}_{k_i}),
\label{likelihood_fn_2}
\end{align}
where $u_i$ is the solution of one of the three differential equations (\ref{lewis_terr_2Dpde_ss}), (\ref{lewis_terr_2Dpde_sta}) or (\ref{lewis_terr_2Dpde_pa}).  Using the AIC model selection techniques outlined in the previous selection, equation (\ref{lewis_terr_2Dpde_pa}) turns out to fit the data significantly better than the other two models, suggesting that movements towards available prey, together with conspecific avoidance (CA), explain the territorial patterns better than either avoidance of steep terrain plus CA or just CA on its own \cite{moorcroftetal2006} (figure \ref{MLMplot}).  Notice that AIC penalises the number of model parameters.  Therefore it is important that we use the non-dimensional forms of the models, which have the minimum number of parameters needed to specify the model.

While such reaction-diffusion approaches have been successful in both describing territorial patterns and inferring behavioural features, the above methods do not give sufficient evidence to conclude that the best fit parameters values of $\beta, m, \alpha_z$ accurately reflect the underlying mechanisms.  First, reaction diffusion approximations can sometimes fail to describe the patterns that arise from underlying individual-level descriptions, especially when interactions are rare, as is the case with territorial behaviour.  Second, though the models are built from one level of description, the movement and interaction mechanisms, they are fitted to data on another level, the space use patterns.  From a logical perspective, it is not necessarily true that a good fit to space use implies an accurate description of the underlying movement and interaction mechanisms.  

\begin{figure}[ht!]
\includegraphics[width=\columnwidth]{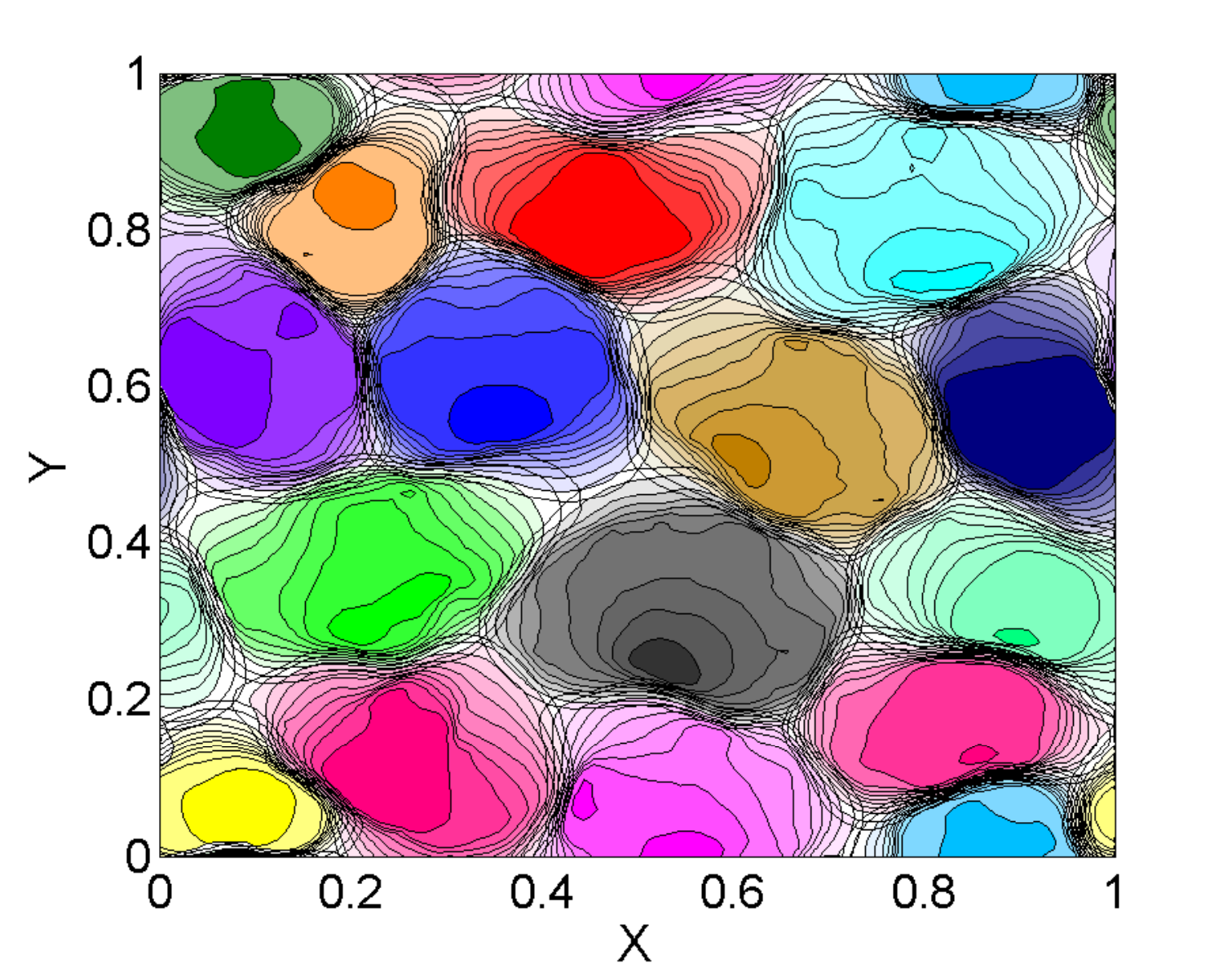}
\caption{\small {\bf Output from an individual-level model of territory formation.}  Contours show the utilisation distribution of various animal positions from an individual based model of territory formation with periodic boundary conditions.  Each colour denotes a different animal's territory (based on \cite{GPH1}).}
\label{2dplot}
\end{figure}

The first issue was addressed recently by building a model of territoriality by directly simulating individual-level movement and interaction processes \cite{GPH1}.  Territorial patterns emerged that well-fitted long-term data on fox movements and could be used to infer information about the longevity of scent cues that accurately replicated field observations \cite{PHG3} (figure \ref{2dplot}).  One striking difference between this approach and that of \cite{moorcroftlewis2006} is that no anchoring den site is necessary to see territorial patterns emerge.  This causes the territories to continually move, never settling to a non-trivial steady state, which means the approach of analysing steady-state ODEs as in equations (\ref{lewis_terr_2Dpde_ss}), (\ref{lewis_terr_2Dpde_sta}) and (\ref{lewis_terr_2Dpde_pa}) is no longer usable.

\begin{figure}[ht!]
\includegraphics[width=\columnwidth]{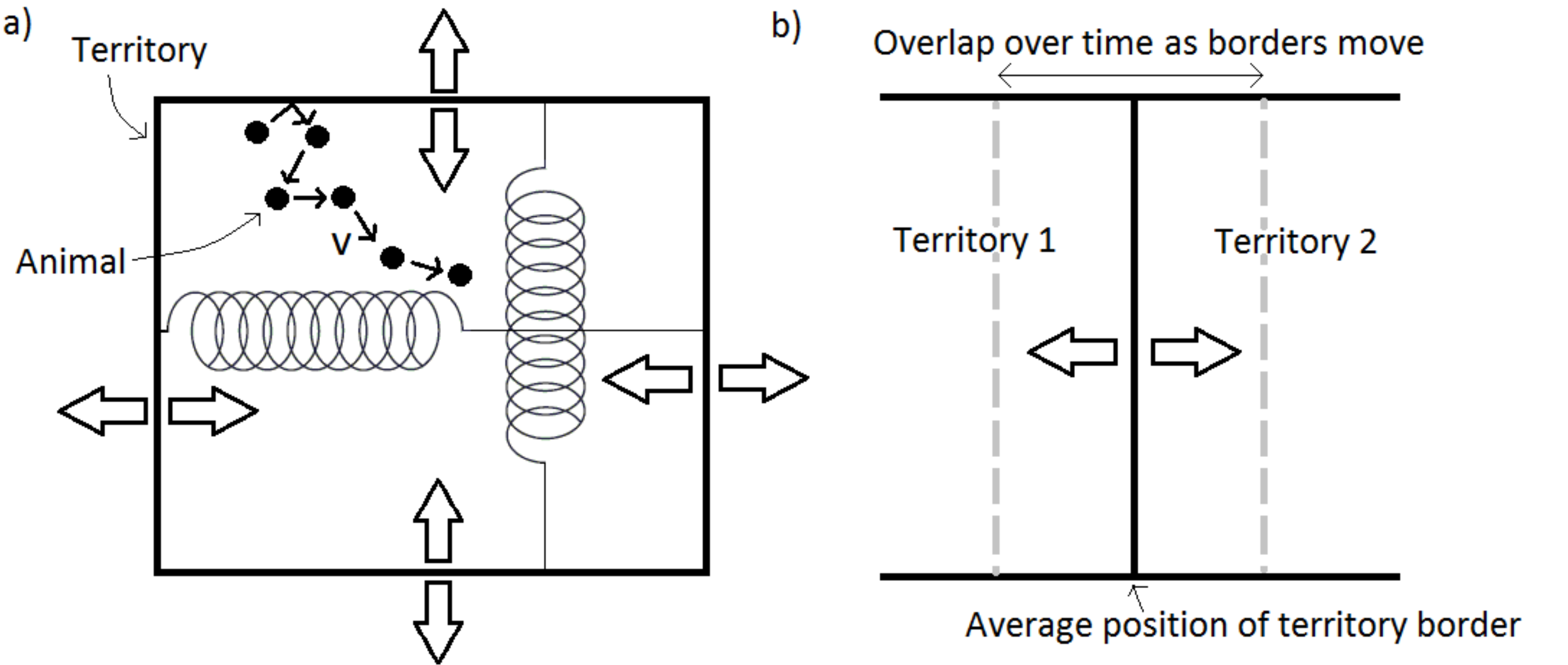}
\caption{\small {\bf Schematic of an analytic model of animal movement within a dynamic territory.}  Territory borders move in a subdiffusive fashion owing to territorial exclusion.  The process that keeps the territories at this average width is represented by two springs, one vertical and one horizontal.  The animal, represented by a filled circle in panel (a), moves as a random walker, with a certain amount of correlation between the direction of successive steps.  Panel (a) represents this setup, while panel (b) demonstrates how overlaps between adjacent home ranges arise from this model as animal positions are measured over time.  In panel (b), the mean position of the territory border is represented by the solid black line, while the average extent of the movement of this border to the left and right is represented by the dashed grey lines (based on \cite{PHG3}).}
\label{territory_model_expl}
\end{figure}

Instead, the authors developed semi-analytic techniques for describing the animal's movements based on trends observed in the movement of simulated territories \cite{GPH2, GPH3, PHG2}.  They noticed that the territories move in a subdiffusive fashion, as predicted by the theory of exclusion processes, and that the generalised diffusion constant of the territory decays exponentially with the dimensionless product $D\rho T$, where $D$ is the intrinsic diffusion constant of the animal, $\rho$ is the population density and $T$ is the scent-mark longevity.  This enabled them to build an analytic model of animal movement within territory borders (figure \ref{territory_model_expl}) which could be fitted to movement data.  However, as yet this technique is unable to infer the territorial interaction processes without using simulation analysis, though see \cite{PHG1} for some first steps towards rectifying this.

The second issue can be addressed by extending the program of building movement models described in the previous section `Getting the model right' to include territorial, as well as environmental, interactions.  Rather than constructing one function for all animals, as in equations (\ref{aic_models}), this approach requires constructing different functions for different animals or packs, then coupling them together via the territorial interactions (figure \ref{cssf}).  The generic form of such a model is
\begin{align}
p_i^{t,\tau}({\bf x}|{\bf y}) \propto \phi_i({\bf x}|{\bf y}){\mathcal W}_i({\bf x},{\bf y},{\mathcal E}){\mathcal C}_i({\bf x},{\bf y},{\mathcal P}_i^t),
\label{cssf_general}
\end{align}
where $\phi_i({\bf x}|{\bf y})$ denotes the intrinsic movement of a single animal $i$ on its own, e.g. equation (\ref{step_length_dist}), ${\mathcal W}_i({\bf x},{\bf y},{\mathcal E})$ represents interactions with the environment ${\mathcal E}$, as in equations (\ref{aic_models}), and ${\mathcal C}_i({\bf x},{\bf y},{\mathcal P}_i^t)$ is a coupling term representing the interactions between the various animals, such as territorial avoidance.  The term ${\mathcal P}_i^t$ contains the information about the population required to describe these interactions \cite{PML}.

By using methods identical to those in the previous section, one can test candidate models of territorial interaction processes and parameterise an individual-level movement and interaction model.  In the same way as we moved from equation (\ref{step_length_dist}) to (\ref{me}), we can use equation (\ref{cssf_general}) to construct a master equation to derive predicted spatial patterns.  This can either be solved numerically, or a continuous-time PDE limit may be found, which might give some analytic insight.

Before we draw general conclusions, it is interesting to note that models of the sort we describe here also have been applied to human populations. In ground-breaking work, Andrea Bertozzi and colleagues have started to understand the mathematics of crime \cite{short2010}.  Their approach to mapping Los Angeles gang territories was to fit a modified version of equation terrain-taxis model \ref{lewis_terr_2Dpde_sta}, that includes structures impeding gang movement, such as freeways and rivers, instead of terrain elevation.  This was fitted to an extensive Los Angeles database on gang reports, and the analysis provided new insights regarding gang interactions \cite{smith2012}.

\section*{Conclusion and future directions.}

The focus of this paper has been to demonstrate how to derive movement and interaction mechanisms from animal location data, and use these to construct models of territorial patterns.  We have given a brief, pedagogical overview of the various techniques used so far to attack this problem, which we hope will leave the reader in a position to begin using them, together with information about what to read to obtain a deeper and more thorough understanding.

While most current approaches build models of territory formation from plausible movement and interaction mechanisms, then validate the model by fitting it to data, the recent attempts to derive spatial patterns from ready-parametrised movement-and-interaction models give a more conservative approach, which is likely to be more accurate at uncovering the actual mechanisms used by animals.  This is vitally important in predicting the effect of future environmental changes on animal populations in a quantitatively as well as qualitatively accurate way.  

Currently this approach is in its infancy.  The challenge for the future is to build mathematical theory that details the best ways to use equations such as (\ref{cssf_general}) to derive spatial patterns.  The simplest way is numerical derivation.  However, it is more mathematically pleasing, and may save computational effort, to derive a theory of PDEs for the so-called {\it coupled step selection functions} of equation (\ref{cssf_general}) \cite{PML}.  

There are various approaches to deriving PDEs from individual-level descriptions, reviewed in \cite{hillenpainter2013} in a biological context.  Different limits of the master equation may uncover different biological aspects of the patterns.  It would be an important future advancement to see which limits give rise to accurate territorial structures.

Another approach, used more in the physics literature, are van Kampen approximations of Markovian processes.  Recent work by Alan McKane and others has shown that, when biological models exhibit behaviours quite different from those of mean field models, van Kampen approximations often do a better job \cite{mckanenewman2004, mckanenewman2005, alonsoetal2007}.  Since these approximations result in the mean field description in certain limits, they can often be used to tease out the reasons why and how mean field approaches may fail.

It is natural to ask whether territorial animals might try to modify their behaviour so as to gain an advantage over their neighbours.  Here neighbouring packs could effectively play a {\em spatial game} where each tries to maximize its fitness via increased resource consumption arising from territorial expansion while, at the same time, attempting to minimize losses incurred through territorial altercations.  To play the game, the packs should thus be able to modulate their spatial movement behaviours, described in the partial differential equation models, in pursuit of enhanced fitness.  Preliminary work applying the theory of differential games to one-dimensional territorial pattern formation has shown how certain movement behaviours are stable from an evolutionary perspective while others are not \cite{hamelin2010}.  One fascinating aspect of this analysis has been it's ability to explain the spontaneous emergence of {\em buffer zones}, where neither pack is found, between wolf territories as the outcome of an evolutionarily stable strategy.  Such buffer zones have observed in nature, and have been studied in detail for wolf populations in northeastern Minnesota \cite{mech1977}.


In conclusion, while we have gone a significant way to understanding the mathematics behind territory formation, much work needs to be done.  In this era of rapid ecological change, predictive ecology is becoming an increasingly important subject.  With ever-changing ecosystems, understanding the mechanisms behind observed spatial patterns is vital for such predictions to be possible.  We are on the first steps of a journey towards making ecology quantitatively predictive.   But it is one that cannot be tackled by a small number of scientists.  We hope that this paper has helped you understand this area and its importance, and perhaps encouraged you to join us in this endeavour.

\section*{Acknowledgment.}
This study was partly funded by NSERC Discovery and Acceleration grants (MAL, JRP).  MAL also gratefully acknowledges a Canada Research Chair and a Killam Research Fellowship.  We are grateful to members of the Lewis Research Group for helpful discussions.

\vfill\eject

\end{document}